# Pressure Wave Detection and Localization in Deployed Underground Fiber using Coherent Correlation OTDR

Florian Azendorf, André Sandmann, Michael Eiselt

*Advanced Technology, Adtran Networks SE, Märzenquelle 1-3, 98617 Meiningen, Germany*
*florian.azendorf@adtran.com*

**Abstract:** A deployed fiber with in-house and underground sections is interrogated with a coherent correlation OTDR. The origin and propagation speed of a hammer-generated pressure wave in the underground section is detected and acoustic signals are monitored. © 2024 The Author(s)

1. **Introduction**

Utilizing deployed optical fibers for environmental sensing has recently gained a lot of attention in academia [1] and industry [2]. Studies in the literature showcase different applications such as earthquake detection [1], traffic monitoring [2], and critical infrastructure monitoring [3], using the deployed fiber network infrastructure as a sensor. These sensors are widely deployed by operators of telecom networks, and can be utilized jointly for communication and sensing. The overall objective of distributed sensing is to spatially resolve environmental effects acting on the fiber, leading to amplitude, phase, and polarization changes of the propagating optical signal. Different approaches can be implemented to evaluate the fluctuations of these parameters by using either the optical signal in transmission or reflection. The most sensitive parameter is the phase of the propagating light wave, and thus, we investigate the phase variations of the backscattered light using a coherent correlation optical time domain reflectometer (CC-OTDR). It provides spatially resolved backscattering information, enabling the monitoring of critical infrastructure and localizing potential threats to predict and prevent damage. In this work, we present the results of a field trial, interrogating a deployed underground standard single-mode fiber with a total length of 8.1 km connected to 280 meters of in-house cabling in our building. We measured acoustic signals in the in-house section caused by air conditioning systems and identified other acoustic noise sources inside the building by evaluating the optical phase. Furthermore, pressure waves generated with a hammer on the company ground nearby the deployed underground fiber were detected, and the event location and propagation speed of the pressure wave in soil were determined.

2. **Experimental setup**

Figure 1 shows the experimental setup of the CC-OTDR connected to the deployed fiber, starting at the Adtran office and ending in downtown Meiningen, Germany. A continuous wave (CW) signal of a highly coherent laser with a Lorentzian linewidth of less than 100 Hz is fed into a polarization-maintaining coupler (PMC), which equally splits the power of the optical signal. One part is sent to the local oscillator (LO) input of the coherent receiver, while the other part is fed into a Mach-Zehnder modulator (MZM). The probe modulation signal is generated by an arbitrary waveform generator (AWG) and it is amplified accordingly. Binary phase shift keying is used as modulation format for the probe sequence, and the baud rate is set to 125 MBaud, leading to a spatial resolution of 80 cm. The probe signal consists of an 8191-bit pseudo-random binary sequence (PRBS) plus a trailing "-1" symbol to obtain a balanced sequence, and a zero padding consisting of 21250 symbols, yielding a frame duration of 235.5 µs. Therefore, the test fiber is probed at a sampling rate of 4.25 kHz. The probe signal is amplified with an erbium-doped fiber amplifier (EDFA). Optical filtering of the probe signal is not applied, since no other signals are propagating in the fiber and the amplified spontaneous emission noise of the amplifier is filtered in the receiver by electrical low pass filters (LPF). Subsequently, the probe signal is coupled to the standard single-mode fiber of 8.4 km length via an optical circulator. The first 600 m of the installation of the fiber are known, where the first 280 meters are in-house cabling and the subsequent 320 meters are a deployed underground fiber close to our building. Hence, all the events are measured and generated in the first 600 m because they can be verified by appropriate means. Backscattered and reflected signals are transferred to the signal input of the coherent receiver, extracting amplitude, phase and polarization via homodyne detection. The four extracted components are sampled with a real-time oscilloscope with a sampling rate of 625 MS/s and stored for offline signal processing. The recorded signals are cross-correlated with the transmitted PRBS sequence. As compared to transmitting a long single pulse into the sensor fiber, transmitting a pulse sequence and perfoming a cross-correlation improves the spatial resolution while maintaining the same sensing reach. The return loss trace in Fig. 1 shows parts of the in-house cabling plus the initial section of the outdoor installation. Peaks with amplitudes in the order of -45 dB down to -60 dB are caused by reflections at connector pairs, whereas all the other peaks correspond to the Rayleigh backscattering pattern. The blue dots are backscattering maxima being selected by a peak search algorithm, which are used to determine the phase evolution at various positions along the fiber.

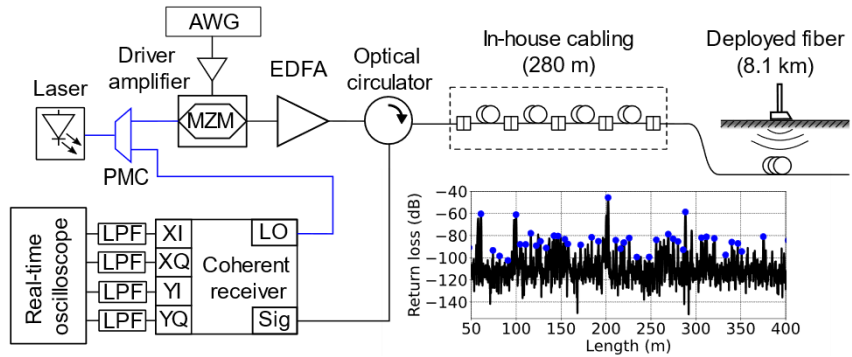

Fig. 1. Coherent correlation OTDR setup interrogating the in-house and deployed underground fiber sections.

## 3. Results

The in-house cabling section is partly deployed in cable channels, which are attached to the ceiling. Other parts are installed under the elevated floor. Hence, the fiber experiences acoustic noise caused by air conditioning systems and other acoustic noise sources in the laboratory environment. In Fig. 2a, the optical return loss, i.e., the characteristic fiber fingerprint along the fiber, is shown as a function of time. The four vertical red lines correspond to high reflections from connector pairs, which stay constant over time. Furthermore, the orange lines are maxima in the Rayleigh back-scattering fingerprint. These lines are almost stable over time where no acoustic signal is present, e.g. at 115 m. With an acoustic signal present, the locally backscattered power varies over time, for instance, as shown in the range between 80 m and 100 m. While the variation of the backscattered power is less sensitive towards external vibrations, the propagation phase can be used as a more sensitive parameter. Fig. 2b shows the resulting waterfall plot, which is the phase change over time as a function of fiber position. For instance, at 74 m to 82 m (I), the phase changes are caused by the air conditioner in the laboratory, which causes vibrations at approximately 50 Hz. This value was verified by measuring the vibration frequency with the accelerometer of a mobile phone. The phase signals showed additional acoustic events caused by different noise sources, e.g., 98 Hz from a server fan. At 116 m to 122 m (II), a weak phase variation with 50 Hz can be noticed. This variation comes from another air conditioning system in a different department in the building. Additionally, the short impulses between 327 m and 346 m (III) stem from a pressure wave applied by a hammer to the ground outside the building. Two strikes were applied to the ground during the record time. It is worth noting that the fiber is deployed in the ground at approximately 0.5 m depth. While it was not feasible to extract the events by analyzing the amplitude of the recorded signals by evaluating the phase, the individual events can be clearly identified and localized.

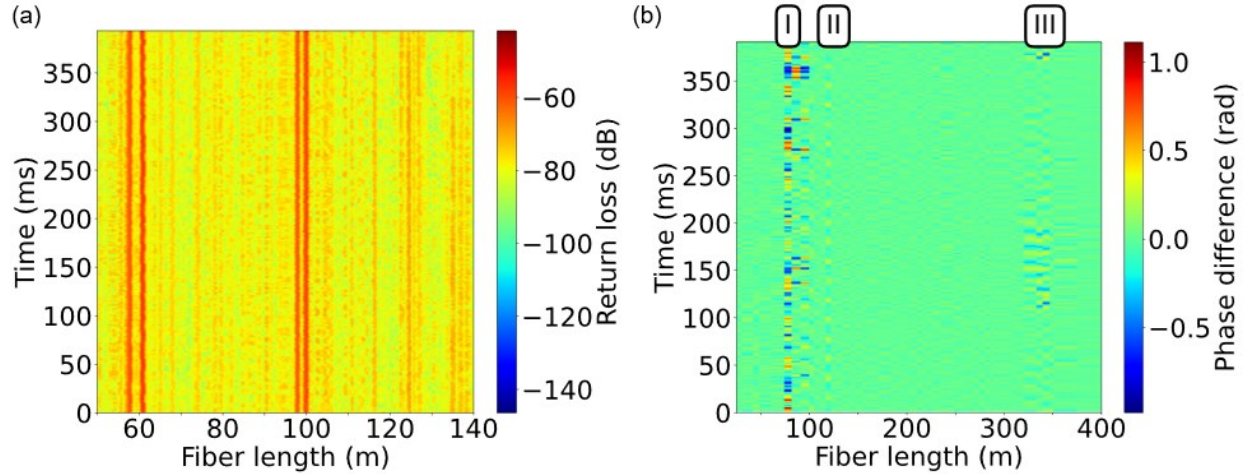

Fig. 2. (a) Time evolution of the fingerprint waterfall, (b) time evolution of phase along the test fiber.

The generated hammer impact on the soil leads to a pressure wave traveling through the ground, affecting the deployed fiber cable. Different cable sections are affected at different times. In order to investigate this, the phase signals of different fiber sections are analyzed. The process is highlighted in Fig. 3 for the y-polarization of the backscattered signal. Firstly, peaks of the fiber fingerprint are selected, see Fig. 3a. Subsequently, the phase differences between the peaks are calculated for each test sequence transmitted into the fiber, i.e., repeated every 235.5 µs. The time evolution of the phase differences is shown in Fig. 3b for six fiber sections. The impact of the pressure waves leads to strain changes in a fiber section, resulting in a high phase variation. Two pressure waves from two consecutive hammer impacts are visible in the graph, one starting at 105 ms and the other at 370 ms. The phase variations show a duration of approximately 100 ms. In the next step, the arrival times of the pressure wave in the individual fiber sections are extracted. The arrival times and the fiber section center locations are mapped in Fig. 3c. The process is repeated for the x-polarization to reduce the fading impact. The pressure wave is assumed to propagate linearly along the fiber length. A corresponding linear least squares fit is performed with a root-mean-square error of 1.6 ms. It shows that the location of the hammer impact is at 336.8 meters. The speed of the pressure wave traveling through the soil is 433 m/s, measured in dry weather conditions at 22°C air temperature.

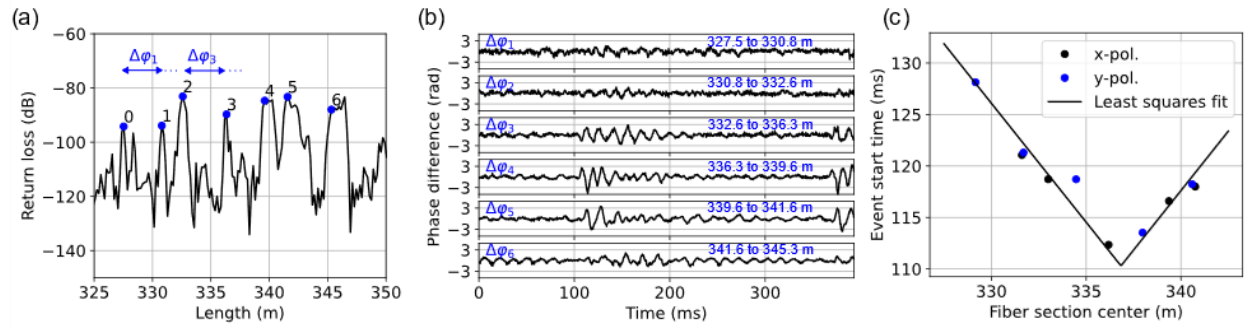

Fig. 3. (a) Fingerprint peak selection, (b) peak phase differences of the y-polarization showing the pressure wave impact at different fiber sections, (c) extracted pressure wave arrival time on different fiber section centers, and pressure wave model fit.

4. **Conclusion**

In this study, we successfully conducted a field trial using a deployed standard single-mode fiber connected to our company building with a coherent correlation OTDR interrogator. We were able to measure phase variations caused by the in-house air conditioning system with 50 Hz and other acoustic noise sources at different sections along the fiber. In addition, pressure waves generated with a hammer near a deployed underground fiber section could be located, and the velocity of the pressure wave in soil was determined to be 433 m/s.

5. **Acknowledgements**

This work has received funding from the Horizon Europe Framework Programme under grant agreement No 101093015 (SoFiN Project) and was partially funded by the German Federal Ministry of Education and Research in the framework of the RUBIN project Quantifisens (Project ID 03RU1U071D).